\documentclass{amsart}

\usepackage[english]{babel}
\usepackage{hyperref}
\usepackage{fancyhdr}
\newcommand{\abbr}[1]{{\sc\lowercase{#1}}}
\newcommand{\osn}{\oldstylenums}

\title{Open Source software and peer review}
\author{Sanatan Rai}
\email{sanat@stanford.edu}
\address{Department of Management Science and Engineering,%
Stanford University, Ca \osn{94305}}

\begin{document}
\begin{abstract}
We compare open source software development to peer review in
academia.
\end{abstract}\maketitle
The debate between Open Source (\abbr{os}) and closed source proprietary
software has been going on for some time. Proponents
are usually passionate in their support and sometimes virulent in their
criticisms. 

In this article I compare the Open Source development methodology
to the system of peer review in academia. 
My hope is to focus attention on the 
practices that can be implemented and  disseminated, rather
than criticise this or that firm for producing poor quality software.

\section*{Two styles of software development}\label{sec:soft}
Let us first examine two famous \abbr{os} software: the {\TeX} typesetting
system and the Linux kernel. 

As is perhaps well known, {\TeX} was written by Don Knuth in the '\osn{70}s
because he was fed up of the poor quality of the proofs from his publisher.
He learnt the fundamentals of typesetting and font design, and produced
{\TeX} and MetaFont. 

At this point, there are no known bugs in either programme. Knuth promised
a small monetary remuneration for each bug found, and not surprisingly the
two programmes are supremely robust  that do exactly what 
their author intended them to do.

The entire source for {\TeX} can be downloaded from any of the 
\abbr{ctan} archives. Knuth discusses the  code in detail in 
{\sl {\TeX}: The program} \cite{texbook}. It is likely that
better versions of many of the algorithms used in {\TeX}
are known today. However, as yet there exists no programme or
package that provides an implementation as robust and versatile
as {\TeX}. Commercial typesetting packages such as Adobe's Acrobat
Writer produce wonderful documents with great ease, however, they 
cannot beat {\TeX} at what it was designed to do. In particular,
for mathematical typesetting there is nothing that comes even
close to {\TeX}.

The Linux kernel was developed initially by Linus Torvalds
from the {\tt minix} kernel designed by Andrew S Tanenbaum.
Torvalds initially developed the kernel for his own use, but its
popularity spread rapidly and within a couple of years it had
become a major volunteer project. 

The Linux operating system, which is built around the Linux
kernel, is now widely used. It is extremely popular in universities
and in the Unix community. It has become something of a showcase
for the \abbr{os} community. 

As a programme it is not bug-free. In fact it is 
under active development, and 
major versions are brought out on a frequent basis. Major
criticism of the kernel has centred around its not satisfying
the definition of `a modern operating system kernel', for instance
being \emph{monolithic} as opposed to being a \emph{microkernel}, 
ie merely an arbiter for the interactions of various processes.

In any case, the Linux kernel has a solid presence in the world
of computers. Its code is available free and in its entirety. Whereas
many volunteers work on various bits of the kernel, a select few
ultimately decide which of those percolate to the released
version. The code is rigorously tested, and bugs are found quickly
and usually eliminated. 

In contrast to these programmes, we have the world of closed source, 
proprietary
software. The contrast is primarily in the culture and the mechanism
of development, though \abbr{os} advocates gleefully point to the 
poor quality of some very popular commercial software. 

It should be unfair to pretend that all commercial software is bad.
Some products, such as Adobe Writer and Rational's Purify have a 
wide following and are held in high regard by their users. Therefore,
the problem is not with the source code being closed to outside
scrutiny per se.

Typical complaints
against bad software are:
\begin{enumerate}
\item there is a big gap between advertised or perceived capability
and actual capability,
\item the vendor did not fix common bugs even on repeated requests,
\item there are \emph{obvious} bugs, ie bugs that could have been eliminated
by minimal testing on the part of the vendor, and
\item poor reliability.
\end{enumerate}

There are thus two main complaints: poor design and implementation, and the 
irresponsibility of the vendor in dealing with bug reports and customer 
problems. 

Vendor irresponsibility depends on the social culture within the company
and its market strength. We shall not discuss it here. 

The design and the implementation on the other hand are the major concern.
These depend primarily on the quality of developers hired by the vendor
and their attitude to the product. This is something that the customer 
cannot control directly. 

In this enviroment, the only instances when good design and robust 
implementation are enforced
is when the vendor interacts closely with the customers, and respects
and values their suggestions and needs. What this does is to make 
the customer a direct part of the testing, debugging and extension
cycle. This is the best that can be done in the absence of open 
code scrutiny. 

Therefore, in case of the closed source vendor, the quality of the
product depends very substantially on the ability and integrity of
the individual developers. There is no way of ensuring good design and 
implementation. This is the biggest weakness of the closed source system.

There is good reason to believe that closed source 
companies regard this situation as being very convenient. It permits
them to charge premium prices for bug fixes and upgrades. It allows
them to get away with not doing a good job. Considering that software
downtime costs the world economy billions of dollars per annum, this
is a serious situation. 

\section*{Publications and peer review}\label{sec:pub}
The standard paradigm for publication of new work in academia is
submission to a refereed journal. An academic writes a paper, he then
sends it to a journal in his field. The journal then has the article
reviewed by experts who then inform the journal whether the paper is
fit for publications. An article is regarded fit for publication in
a research journal if it:
\begin{enumerate}
\item presents substantially new and original results, or
\item presents a substantially novel and useful interpretation or
application of known results, or
\item is of an expository nature, and provides a summary or survey
of the latest developments in a field.
\end{enumerate}
So the purpose of peer review is to ensure the quality of published
papers. The system as it exists has been criticised in recent times,
there has been much debate over electronic versus print journals. 
It is felt by many that the system in practice is not very effective.
However, no one disputes the necessity of peer review. In fact the biggest
critics of the current system lament that in practice, adequate \emph{review}
is not provided by the system \cite{not}

The quality of papers in the best journals is witness to the efficacy of
this system. Of course, much depends on the integrity and ability
of the individual authors. Since every academic is not of the calibre
of Grothendieck (who incidentally did not care to publish very much) or
Feynman or Knuth, the academic world depends on peer review to maintain
the quality of papers and to detect errors. In fact even great
mathematicians make mistakes: a famous recent example 
is Andrew Wiles' proof of Fermat's last theorem. His initial proof
had an error, which was caught by the referees. He then
emended the error, and the new corrected proof was finally
published in the \emph{Annals}, about a year after the initial submission
\cite{wiles}. 

\section*{Software and peer review}\label{sec:peer}
The greatest strength of open source software is the peer review.
The free and unrestricted availability of the source code
leads to open scrutiny and debate. At the very minimum, it expands
the tester base. The result is that the more popular or
crucial an open source project is, the more likely it is to be
scrutinised, and thus such projects tend to be robust, well
designed, and have fewer bugs. Bugs are also more likely to be detected
and eliminated. 

Recently there has been a major push by some of the bigger companies 
based on the closed source model, to stifle testing and review. 
Not surprisingly, the first victims were researchers who pointed out
major flaws in the design. Rather than being thanked they were 
threatened with arrest in one case, and actually arrested in another.

This kind behaviour is pernicious. It rewards poor design and practice.
It will lead to the deployment of unreliable and insecure systems
that will be easy pray for the maliciously inclined. At the very
minimum they are going to cost the world economies huge losses.

\section*{Conclusion}\label{sec:conc}
In this article I have equated the open source system to the system
of peer review in academia. The advantages of the latter are clear
to all, and the same advantages can be found in the case of open source
software development. 

The quality of closed source software is too dependent on the ability
of the individual programmers. The software industry should be aiming
to produce better and more robust software. It is perverse to think that
the short term gain that companies make by dragging their feet over
bug correction and testing, compensates for the huge losses
made by the economy as a whole.
\bibliographystyle{amsalpha}
\bibliography{OpenSourceAndPeerReview}

\end{document}